\documentstyle[12pt,aps]{revtex} 

\oddsidemargin=\evensidemargin
\input epsf 

\title{Pionic decay of a possible $d^{\prime}$ dibaryon 
       and the short-range NN interaction\thanks{supported 
                          with DFG Grant No. Fa-67/20-1 and 
                               DFG postdoctoral fellowship Wa1147/1-1}}
\begin{document}
\maketitle
\begin{center}
I.T.\ Obukhovsky$^{1,2}$, K.\ Itonaga$^{2,3}$, 
Georg Wagner$^2$, A.J.\ Buchmann$^2$ \\and Amand Faessler$^2$
\\[0.15cm] 
$^1${\footnotesize{\it Institute of Nuclear Physics, Moscow State
University, 119899 Moscow, Russia}}
\\
$^2${\footnotesize{\it Institute for Theoretical Physics, 
University of T\"ubingen, Auf der Morgenstelle 14,\\
D-72076 T\"{u}bingen, Germany}}
\\
$^3${\footnotesize{\it Laboratory of Physics, Miyazaki Medical College
Kiyotake, Miyazaki 889-16, Japan}}
\end{center}


\begin{abstract}

We study the pionic decay of a possible dibaryon 
$d^{\prime}\to N+N+\pi$ in the microscopic quark shell model. 
The initial $d^{\prime}$ dibaryon wave function (J$^P$=0$^-$, T=0) 
consists of one 
1$\hbar\omega$ six-quark shell-model $s^5p[51]_X$ configuration.
The most important final six-quark configurations
 $s^6[6]_X$, $s^4p^2[42]_X$ and $(s^4p^2-s^52s)[6]_X$ are properly  
projected onto the NN channel.        
The final state NN interaction is investigated by means of two 
 phase-equivalent - but off-shell different - potential models.
We demonstrate that the decay width $\Gamma_{\rm{d'}}$ depends strongly on
the short-range behavior of the NN wave function.
In addition, the width $\Gamma_{\rm{d'}}$ is very sensitive to the 
mass and size of the $d^{\prime}$ dibaryon. 
 For dibaryon masses slightly above the experimentally suggested value
$M_{\rm{d'}}$=2.065 GeV, we obtain a pionic decay width of 
$\Gamma_{\rm{d'}}\approx$ 0.18--0.32 MeV close to the 
experimental value $\Gamma_{\rm{d'}}\approx$ 0.5 MeV.

\end{abstract}

\newpage
\section{Introduction}

During the last decade much attention has been devoted to 
theoretical and experimental investigations
of the pionic double charge exchange
($\pi$DCX) process on nuclei. 
Because this reaction  $\pi^++(A,Z)\to (A,Z+2)+\pi^-$ involves (at least) two 
nucleons in the nucleus, the $\pi$DCX cross section depends     
sensitively on short-range NN-correlations in nuclei. 
Therefore, it provides a good testing ground for
the nucleon-nucleon interaction at short range.
Experiments on different nuclear targets have unambiguously 
confirmed the existence of a narrow resonance-like structure in the $\pi$DCX 
cross-section at small incident pion energies 
$T_{\pi}\approx50$ MeV \cite{bil91}. The position of this peak turned out 
to be largely independent of the studied nucleus. 
The height and width of this peak could 
not be explained by standard calculations based on the 
two-step process \cite{kam} ($n+n+\pi^+\to n+p+\pi^0\to p+p+\pi^-$). 
So far, these data could only be explained with the assumption 
of a non-nucleonic reaction mechanism \cite{bil91,mart91} proceeding via an
intermediate dibaryon resonance, henceforth called $d^{\prime}$.
The quantum numbers of the $d^{\prime}$ dibaryon candidate were
determined as J$^P$=0$^-$, T=0, and its free mass and hadronic 
decay width were suggested to be $M_{\rm{d'}}$=2.065 GeV and
$\Gamma_{\rm{d'}}\simeq$ 
0.5 MeV\footnote{This value is uncertain by a factor of two \cite{Clem95}.}.
More than a decade ago
Mulders {\sl et al.\ } \cite{muld} predicted a dibaryon 
resonance with quantum numbers
$J^P$=$0^-$, T=0 and a mass $M\approx 2100$ MeV within the MIT bag model.
Recently, this dibaryon candidate has been investigated in a series 
of works \cite{wag95,buch95,iton} within 
the T\"{u}bingen chiral constituent quark model.
These works emphasize the crucial role of the confinement mechanism for the 
existence of the $d^{\prime}$.

The quantum numbers J$^P$=0$^-$, T=0 of the $d^{\prime}$ resonance prevent the
decay into two nucleons and the only allowed hadronic
decay channel of the $d^{\prime}$ is the three-body decay into a 
$\pi NN$ system with S-waves in each particle-pair \cite{bil91,mart91}. 
Because the
$d^{\prime}$ mass $M_{d^{\prime}}$ is only $\approx 50$ MeV above the $\pi NN$
threshold, the $d^{\prime}$ decay width $\Gamma_{d^{\prime}}$ should be anomalously
small owing to a very small phase volume of three-particle final states.
We recall that the currently available experimental evidence
of dibaryon excitations in nuclei is very limited \cite{seth}.
This is due to very large N-N decay widths of most dibaryon
resonances, which renders them undetectable on
the background of other hadronic processes at intermediate energy.
At present, the experimental
 evidence for narrow dibaryons
  is reduced to a single candidate, the  $d^{\prime}(2065)$.
In contrast to the deuteron, which consists of two on the average widely
separated nucleons, there are indications \cite{wag95,buch95}, 
that the $d^{\prime}$ is a rather pure, compound six-quark system.
Therefore, the dynamics of its hadronic decay into the $\pi$NN system
should be sensitive to the overlap region of the two outgoing nucleons;
a situation that is ideal for understanding
the role of quark degrees of freedom in the short-range
nucleon-nucleon interaction (see e.g.\ Ref.\ \cite{gold95} and references 
therein).

\begin{figure}[h]
  {\epsffile{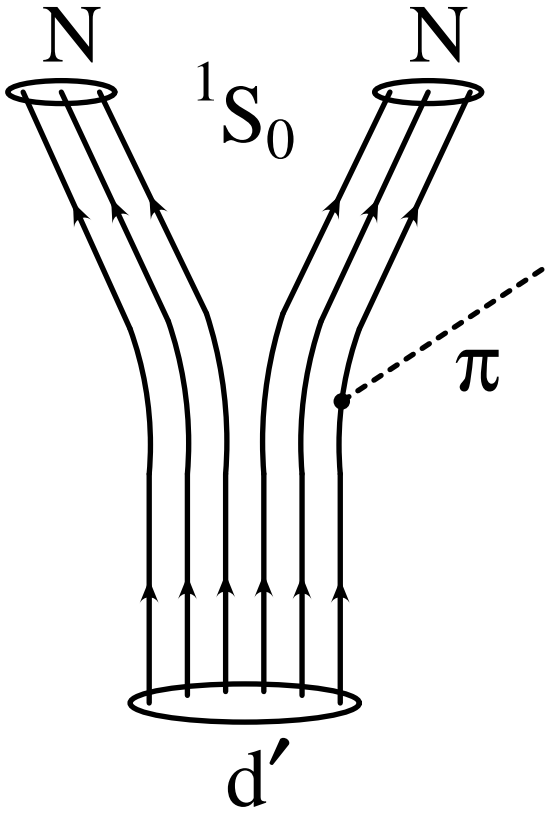}}
\caption[Quark]{Quark line diagram of the pionic dibaryon decay. The elementary
         pion is produced on a single quark, leaving the remaining six quarks
         in a relative $^1S_0$ nucleon-nucleon scattering state.} 
\label{figure:obufig1} 
\end{figure}

Starting from this point 
(for alternative approaches see in Refs.\ \cite{kam,gar,sim})
we consider the $d^{\prime}$ decay as a (quark) shell-model transition 
from one 
six-quark configuration to another one by emitting a pion. 
The quark line diagram of the decay is sketched in Figure \ref{figure:obufig1}.
The calculation of the transition matrix elements $d^{\prime}\to N+N+\pi$ 
is similar to the calculation of $\Delta$-isobar-decay matrix elements 
$\Delta\to N+\pi$ (spin and isospin flip of a quark). 
In the case of the $d^{\prime}$ decay only the initial dibaryon state is a 
definite six-quark configuration 
(the lowest shell-model configuration with quantum numbers J$^P$=0$^-$, T=0),
whereas the final state consists of a continuum of NN-states
which have to be projected onto a basis of six-quark configurations
with quantum numbers J$^P$=0$^+$, T=1 of the NN $^1S_0$ wave. 
The main difficulty in comparing the calculated width $\Gamma_{d\prime}$
with experimental data is its sharp dependence 
on the energy gap between $M_{d^{\prime}}$ and the $\pi$NN threshold.
A reliable result on $\Gamma_{d^{\prime}}$ can be obtained only if the exact
mass $M_{d^{\prime}}$ in vacuum is known 
(e.g.\ from electroexcitation of the $d^{\prime}$ on the 
deuteron at large momentum transfers \cite{schep}).
At present, we have only indirect data in the nuclear medium \cite{bil91}. 
Due to the absence of vacuum data, we investigate the problem of the 
$d^{\prime}$ decay width starting from theoretical quark-model results 
\cite{wag95,buch95} for $M_{d^{\prime}}$ 
and the hadronic $d^{\prime}$ size parameter $b_6$.

Our first calculation for $\Gamma_{d^{\prime}}$ was published 
in Ref.\ \cite{iton}. 
The aim of the present work is to improve mainly on three
important effects which were neglected in Ref.\ \cite{iton}:  
a) antisymmetrization of the final NN-state on the 
quark level taking into account
   the effect of quark exchange between the two nucleons at short range, 
b) insertion of a complete basis of final six-quark states including besides 
   the non-excited $s^6$ shell-model state all Pauli-allowed excited 
   configurations $s^4p^2$ and $s^52s$, which have a non-vanishing overlap
   with the final NN-state and can be populated via the emission of the pion
   from the initial $d^{\prime}$ dibaryon, and
c) inclusion of the final state interaction (f.s.i.)
   for the  two-nucleon system. 
   
   
\section{Decay dynamics in terms of quark degrees of freedom}


\subsection{Initial state}
\noindent
As in Ref.\ \cite{iton} we consider only the simplest
six-quark configuration $s^5p[51]_X$ in the initial state (the energetically
lowest J$^P$=0$^-$, T=0 translationally invariant shell model (TISM)
state which satisfies the Pauli exclusion principle).
It has been shown in \cite{wag95,buch95}, that the $d^{\prime}$ wave function may
be considered as a compound six-quark state, for which a single shell model
vector provides an adequate description.
This state vector is defined by
\begin{equation}
\label{d1}
|d^{\prime}\rangle=
|s^5\!p\,(b_6)\,[51]_X,\, [2^3]_C[3^2]_T([2^21^2]_{CT})
[42]_S:[21^4]_{CTS},\,LST\!=\!110,\,J^P\!=0^-\rangle \; .
\end{equation}
The characteristic oscillator parameter in the six-quark wave function $b_6$ 
may for example be determined from the minimisation of the $d^{\prime}$ mass 
for a given microscopic quark-quark Hamiltonian \cite{wag95,buch95}.
The Young schemes $[f_D]$, D=X,C,S,T in orbital, color, spin and isospin space,
as well as for the coupled spaces CT, CTS, are necessary for the unambigious
classification
of shell-model basis vectors in terms of irreducible representations
 (IR) of the following reduction chain for unitary groups:
\begin{eqnarray}
 SU(24)_{XCST}\supset SU(2)_X\times SU(12)_{CST} &\supset&
 SU(2)_X\times SU(6)_{CT}\times SU(2)_S\supset
\nonumber\\
 && SU(2)_X\times SU(3)_C\times SU(2)_T\times SU(2)_S
\label{su}
\end{eqnarray}
The fractional parentage coefficient (f.p.c.) 
technique \cite{obu79,har,chen,obu82}
based on scalar factors (SF) of Clebsch-Gordan coefficients of the
above group \cite{chen,obu82,obu84,so}, sketched in the following section, 
is used for the calculation of matrix
elements and overlap integrals.

\subsection{Transition operator}
The pionic decay width of the $d^{\prime}$ is calculated, as in Ref.\ 
\cite{iton}, assuming a direct coupling of constituent quarks with the
isotriplet of pion fields $\bbox{\phi}$ through the operator
\begin{equation}
\label{o}
\widehat {\cal{O}}_{\pi q}({\bf k})=\frac{f_{\pi q}}{m_{\pi}}
\,\sum_{j=1}^6\, 
(\bbox{\sigma}_j\cdot {\bf k})\, ({\bbox{\tau}}_j\cdot {\bbox{\phi}})
\,\frac{ \exp(-i{\bf k}\cdot ({\bf r}_j-{\bf R}_{c.m.}))}
{\sqrt{2E_{\pi}(2\pi)^3}} \; .
\end{equation}
Here, ${\bf r}_j$, ${\bbox{\sigma}}_j$ and ${\bbox{\tau}}_j$ are coordinate,
spin and isospin of the j-th quark, {\bf k} is the pion momentum in the
c.m.s.\ of the $d^{\prime}$, and $E_{\pi}= \sqrt{m_{\pi}^2+{\bf k}^2}$.
  $f_{\pi q}$ is the $\pi qq$ coupling constant. Its value  
is connected with the $\pi NN$ coupling  
$f_{\pi N}$  (we use $\frac{f_{\pi N}^2}{4\pi}=0.07491$) through the
known relation
$\langle N(123)|\sum_{j=1}^3\sigma_j^{(z)}\tau_j^{(z)}|N(123)\rangle
=\frac{5}{3}\langle N|\sigma_N^{(z)}\tau_N^{(z)}|N\rangle$,
giving $f_{\pi q}=\frac{3}{5}f_{\pi N}$.
Because we neglect isospin breaking effects in this work,
we chose the average pion mass $m_\pi$ = 138 MeV.


\subsection{Final states}
\noindent
In Ref.\ \cite{iton} the wave function of the final NN-state was
antisymmetrized and normalized on the
nucleon level, assuming a plane wave with wave
vector ${\bf q}$ in the relative coordinate ${\bf r}$ between the two nucleons.
The coordinate representation of the nucleon-nucleon state vector
$|\,\Phi_{NN}({\bf q})\rangle$ was then written as
\begin{equation}
\label{pw}
\langle {\bf r}\, |\,\Phi_{NN}({\bf q})\rangle=\Phi_{NN}({\bf q},{\bf r})=
\frac{1}{(2\pi)^{\frac{3}{2}}}\frac{1}{\sqrt{2}}
\left[e^{i{\bf q} \cdot {\bf r}}-
(-1)^{(S+T)}e^{-i{\bf q} \cdot {\bf r}}\right]
\; ; \; S=0, T=1 \, .
\end{equation}
The full wave function of the final state
respected the three-quark cluster nature of the nucleons
\begin{equation}
\label{tpsinn}
\langle {\bf r}\, |\,\widetilde{\Psi}_{NN}({\bf q},123456)\rangle=
\Phi_{NN}({\bf q},{\bf r})\{N(123)N(456)\}_{ST} \; ,
\end{equation}
i.e.\ the nucleon wave function $N(123)$ was given
by translationally-invariant shell-model configurations 
\begin{eqnarray}
\label{n}
N(123) &=&
|s^3\,(b_N)\,[3]_X,[1^3]_C[21]_T([21]_{CT})[21]_S:
[1^3]_{CTS},LST\!=0,\, 1/2,\, 1/2\rangle_{TISM}
\nonumber \\ 
&=& \Phi_N(123) \,\cdot \, 
\chi^{C=0}\cdot\chi^{S=\frac{1}{2}}_{S_z}\cdot\chi^{T=\frac{1}{2}}_{T_z}\; .
\end{eqnarray}
$\chi^{C=0}$, $\chi^{S=\frac{1}{2}}_{S_z}$ and
$\chi^{T=\frac{1}{2}}_{T_z}$ are color-singlet, 
spin and isospin three-quark states.
$\Phi_N(123)$ is the orbital part of the wave function, expressed in terms of
the internal Jacobi coordinates $\bbox{\rho}={\bf r}_1-{\bf r}_2$ and
$\bbox{\lambda}={\bf r}_3- ({\bf r}_1+{\bf r}_2)/2$
\begin{equation}
\label{phin}
\Phi_N(123)=(\sqrt{3}\pi b_N^2)^{-3/2}\exp\left[-\frac{1}{2b_N^2}
\left(\frac{1}{2}\rho^2+\frac{2}{3}\lambda^2\right) \right] \; ,
\end{equation}
with a characteristic nucleon oscillator parameter $b_N$.
This parameter does not have 
to be the same as the harmonic oscillator parameter
$b_6$ for the dibaryon wave function of Eq.\ (\ref{d1}), as it has
been  discussed in Refs.\ \cite{wag95,buch95}.
The relative Jacobi coordinate between the clusters is given by
\begin{equation}
\label{r}
{\bf r}=\frac{{\bf r}_1 + {\bf r}_2 + {\bf r}_3}{3} -
        \frac{{\bf r}_4 + {\bf r}_5 + {\bf r}_6}{3} \; .
\end{equation}

Note that the six-quark final state (\ref{tpsinn})
is antisymmetrized automatically when the 
state vector $|\widetilde\Psi_{NN}({\bf q})\rangle$ is substituted into
the decay matrix element $<\widetilde\Psi_{NN}({\bf q}); \pi\,
|\,\hat {\cal{O}}_{\pi q}|\,d^{\prime}\rangle$, because  the
initial state of Eq.\ (\ref{d1}) is fully antisymmetric. 
However, the
antisymmetrizer-projector $\hat {\cal{A}}$ ($\hat {\cal{A}}^2=\hat {\cal{A}}$)
contained in the initial state
$\hat {\cal{A}} \,|d^{\prime}\rangle=|d^{\prime}\rangle$ reduces considerably 
the normalization of the final state 
(it cuts all non-antisymmetrized parts of the cluster function of 
 Eq.\ (\ref{tpsinn}) which contain about 90\% of the wave function
 -- see below). 
Therefore, it is important to substitute from the beginning a final state wave
function which is normalized (${\cal{N}}$) and antisymmetrized 
($\hat {\cal{A}}$) on the quark level.
\begin{equation}
\label{psinn}
|\Psi_{NN}({\bf q},123456)\rangle = {\cal N}\hat {\cal{A}}
\{\Phi_{NN}({\bf q},{\bf r}) N(123)N(456)\}_{ST},\,\,\,\, ST=01\; .
\end{equation}
The normalization factor ${\cal N}$ is determined by the
standard orthonormalization condition
\begin{equation}
\label{pnorm}
\langle\Psi_{NN}({\bf q}^{\prime})|\Psi_{NN}({\bf q})\rangle=
\langle\Phi_{NN}({\bf q}^{\prime})|\Phi_{NN}({\bf q})\rangle=
\delta^{(3)}({\bf q}^{\prime}-{\bf q}) \; ,
\end{equation}
which leads to
\begin{equation}
\label{norm2}
{\cal N}^{-2} = \frac{\langle\{\Phi_{NN}({\bf q}^{\prime})N(123)N(456)\}_{ST} \,
\vert\, \hat {\cal{A}} \,\vert\, 
\{\Phi_{NN}({\bf q})N(123)N(456)\}_{ST}\rangle } 
{\langle\Phi_{NN}({\bf q}^{\prime})\, |\, \Phi_{NN}({\bf q})\rangle } \; .
\end{equation}
 The antisymmetrizer-projector $\hat {\cal{A}}$ is
\begin{equation}
\label{ant}
\hat {\cal{A}} = \frac{3!3!2}{6!}\left( \;
1-\sum_{i=1}^3\sum_{j=4}^6 P_{ij}^{XCST} \;\right)=
\frac{1}{10} \left(\; 1-9\, P_{36}^{XCST} \;\right) \quad , \quad 
\hat {\cal{A}}^2=\hat {\cal{A}} \; . 
\end{equation}
$P_{ij}^{XCST}$ is the pair-permutation operator for quarks $i$ and $j$
in orbital, color, spin and isospin space.
It is instructive to calculate the normalization factor (\ref{norm2})
algebraically by factorization of the CST and X parts of the pair permutation
$P_{36}^{XCST}=P_{36}^{CST}\, P_{36}^X$. 
The matrix element of $P_{36}^{CST}$ between
two NN-states in the ST=01 (or 10) channel is very well known
(see  e.g.\ \cite{obu84})
\begin{equation}
\label{p36}
\langle\{N(123)N(456)\}_{ST=01}\; |\; P_{36}^{CST}\; |\; 
\{N(123)N(456)\}_{ST=01}\rangle = - \frac{1}{81} \; .
\end{equation}
Inserting this value into Eq.\ (\ref{norm2}) reduces its right-hand-side to
\begin{equation}
\label{normred}
{\cal N}^{-2} = \frac{1}{10} + \frac{1}{90} 
\frac{\langle\Phi_{NN}({\bf q}^{\prime})\Phi_N(123)\Phi_N(456)\; |\; P_{36}^X
\; |\; \Phi_{NN}({\bf q})\Phi_N(123)\Phi_N(456)\rangle} 
{\langle\Phi_{NN}({\bf q}^{\prime})\; |\; \Phi_{NN}({\bf q})\rangle} \; ,
\end{equation}
where $\Phi_N(123)$ is the orbital part 
of the nucleon wave function (\ref{n}) given in Eq.\ (\ref{phin}). 
The numerator in Eq.\ (\ref{normred}) depends on the form
of $\Phi_{NN}({\bf q},{\bf r})$, but for the plane wave in Eq.\ (\ref{pw}) 
(or for any continuum wave function including f.s.i.) 
it is a finite value, i.e.\ it has to be zero
compared with the $\delta$-function in the denominator. 
Therefore, in our case the second term in Eq.\ (\ref{normred}) vanishes 
and we obtain
\begin{equation}
\label{nrm}
{\cal N}=\sqrt{10}
\end{equation}
Note, that the expression $\langle P_{36}^X\rangle\equiv
\langle\,\Phi_{NN}\Phi_N\Phi_N\, |\, P_{36}^X\, |\, \Phi_{NN}\Phi_N\Phi_N\, 
\rangle / \langle\, \Phi_{NN}\, |\, \Phi_{NN}\, \rangle$ receives its
maximal value =1 in the special case of a Gaussian 
$\Phi_{NN}(r)=(2\pi b_N^2/3)^{-3/4} \exp(-3r^2/4b_N^2)$. 
Therefore, for any relative NN wave function $\Phi_{NN}$ we have the
following constraints:
\begin{equation}
\label{cnstr}
0\,\le\langle P_{36}^X\rangle\le 1\; , \mbox{ or } \quad
9\le{\cal N}^2\le 10
\end{equation}
The value (\ref{nrm}) is equal to the usual identity factor
$\sqrt{\frac{6!}{3!3!2}}$  well-known
in nuclear cluster physics (see e.g.\ \cite{nem}). 
It plays an important role for the projection of six-quark
configurations onto baryon-baryon channels \cite{obu91}. 
 From now on we shall
omit the antisymmetrizer $\hat {\cal{A}}$ in front of the  final state in the 
decay matrix element, but the identity factor (\ref{nrm}) may not be omitted:
\begin{equation}
\label{rnm}
\langle d^{\prime} | \widehat {\cal{O}}_{\pi q}({\bf k})|
\Psi_{NN}({\bf q}),\pi\rangle =
\langle d^{\prime} | \widehat {\cal{O}}_{\pi q}({\bf k})\sqrt{10}\hat {\cal{A}}|
\widetilde{\Psi}_{NN}({\bf q}),\pi\rangle =
\sqrt{10}\langle d^{\prime} | \widehat {\cal{O}}_{\pi q}({\bf k})|
\widetilde{\Psi}_{NN}({\bf q}),\pi\rangle
\end{equation}
The inclusion of this factor, due to the antisymmetrization of the
final two-nucleon wave function on the quark level, improves considerably  
the agreement of the results obtained in Ref.\ \cite{iton} with the
experimentally suggested width.


\subsection{Transition amplitude including intermediate states with 
            up to two harmonic oscillator quanta}
\noindent
As in Ref.\ \cite{iton}, we calculate the decay matrix element 
of Eq.\ (\ref{rnm}) by inserting a complete set of six-quark configurations 
with quantum numbers of the final $^1S_0$ two-nucleon state 
(LST=001, J$^P$=0$^+$)
\begin{eqnarray}
&& \langle\Psi_{NN}({\bf q}) , \pi\,
  |\,\widehat {\cal{O}}_{\pi q}({\bf k})\,|\,d^{\prime}\rangle \! = \! 
  \sqrt{10} \! \! \sum_{(n),\{f\}} 
  \langle\Phi_{NN}^{L=0}({\bf q})\{N(123)N(456)\}_{ST\! =\! 01}  | 
  (n,b_6), \{f\} ,LST\!=\! 001 \rangle
\nonumber\\
&& \qquad\; \langle(n,b_6), \{f\} ,LST\!=\! 001 \, |\, 
   \widehat{\cal{O}}_{\pi q}({\bf k}) 
  \, |\, s^5\!p\,(b_6)[51]_X,\,  [2^21^2]_{CT}LST\!=\!110,\,
  J^P\! =\! 0^- \rangle \; .
\label{tme}
\end{eqnarray}
Here, $\{f\}=\{[f_X],\,[f_{CT}]\}$ and (n) defines quark states
with $n$ harmonic oscillator (h.o.) excitation quanta, i.e.\
$(n)=s^{6-n}p^n, s^{6-2m}(2s)^m, (n=2m)$, {\it etc.}; and  
$b_6$ is the h.o.\ parameter for the six-quark system. 
The summation in Eq.\ (\ref{tme}) extends
over a limited set of Young schemes $[f_X]$ and $[f_{CT}]$:
the possible representations of $[f_{CT}]$ in the sum
in Eq.\ (\ref{tme}) are given by the series of inner products of the 
$[2^3]_C$ color and T=1 $[42]_T$ isospin Young schemes
\begin{equation}
\label{ser}
[2^3]_C\circ[42]_T=[42]_{CT}+[321]_{CT}+[2^3]_{CT}+[31^3]_{CT}+[21^4]_{CT} \; .
\end{equation}
Only two spatial Young schemes $[6]_X$ and $[42]_X$ are compatible
with the even-parity (L=0) N-N partial wave.  
 Further constraints follow from the Pauli exclusion principle, i.e.\
$[f_X]\circ [33]_S\circ [f_{CT}] = [1^6]_{XCST}$.
In the case of full spatial symmetry $[6]_X$, only
one color-isospin state  $[2^3]_{CT}$ is allowed, but the
Young scheme $[42]_X$ of the excited shell-model configurations is compatible 
with each state from the inner product given in Eq.\  (\ref{ser}). 
Our choice of a one-body transition (pion-production) operator defined in 
Eq.\ (\ref{o}) further restricts the number of relevant intermediate states. 
The one-particle operator (\ref{o}) can excite
(or de-excite) only one quark of the initial $s^5p$ state.
Therefore, the complete set of states in Eq.\ (\ref{tme}) is reduced to the
configurations $s^6$, $s^4p^2$ and $s^52s$, knowing that higher one-particle 
excitations can be omitted because of a very small overlap with 
the final NN-state.
Summarizing, the following intermediate states
are taken into account in Eq.\ (\ref{tme}):
\begin{itemize}
\item the energetically lowest (n=0) spatially symmetric state
      $s^6[6]_X[2^3]_{CT}$,
\item the excited (n=2) translationally-invariant
      (orthogonalized to the 2S excitation of the six-quark c.m.) state 
      $(s^4p^2-s^52s)$ with identical Young schemes $[6]_X, [2^3]_{CT}$,  and
\item five excited (n=2) states $s^4p^2[42]_X[f_{CT}]$ with CT Young schemes
      from the inner product of Eq.\ (\ref{ser}).
\end{itemize}

It is interesting to note that all these configurations are also 
important for explaining the short-range  nucleon-nucleon interaction.
This was pointed out almost two decades ago \cite{obu79,har,fa82} and 
thereafter discussed in many papers (see e.g. \cite{faess} and references
therein). Now we believe that a possible $d^{\prime}$ dibaryon has much
potential for providing additional information on the innermost 
part of the nucleon-nucleon interaction, i.e. in the region 
where the nucleons overlap.

\subsection{Final state interaction}
\noindent
To take into account the f.s.i.\ for the two outgoing nucleons, 
we consider separable-potential
representations of the N-N interaction, namely the phenomenological
potential of Tabakin \cite{tab}, and the separable model of Ueda et al.\
\cite{ued}, which is equivalent to the OBEP. 
The wave functions of the $^1S_0$ NN final states for the Tabakin potential
are of the form
\begin{eqnarray}
  \Phi^{L=0}_{NN}(q,r) = (2\pi)^{-3/2}\, \cos\!\delta_0 \, 
& & \left\{  
  {\it j}_0(qr)-\tan\!\delta_0\,{\it n}_0(qr)+
  A(q)\frac{ {\it e}^{-\beta r}}{r}   \right.
\nonumber \\
& & + \left. B_1(q)\frac{ {\it e}^{-\alpha r}}{r}\cos\alpha r+
    B_2(q) \frac{ {\it e}^{-\alpha r}}{r}\sin\alpha r \right\} \; ,
\label{fst}
\end{eqnarray}
while the separable potential model of Ueda {\sl et al.\ } leads to the
$^1S_0$ NN wave function  
\begin{equation}
\label{fsu}
\Phi^{L=0}_{NN}(q,r)=(2\pi)^{-3/2}\cos\!\delta_0\,\left\{
{\it j}_0(qr)-\tan\!\delta_0\,{\it n}_0(qr)+\tilde A(q)
\frac{{\it e}^{-\gamma r}}{r}
3-\sum_{n=1}^N\tilde B_n(q)\frac{{\it e}^{-\beta_n r}}{r}\right\} \; .
\end{equation}
Here, $\delta_0(q)$ is the phase shift of NN-scattering
in the $^1S_0$ wave, and the functions A, $\tilde A$, $B_i$ and $\tilde B_i$
depend on the choice of parameters $\alpha$, $\beta$, $\gamma$ and $\beta_i$
for the two models (see the Appendix).

\begin{figure}[h]
  {\epsffile{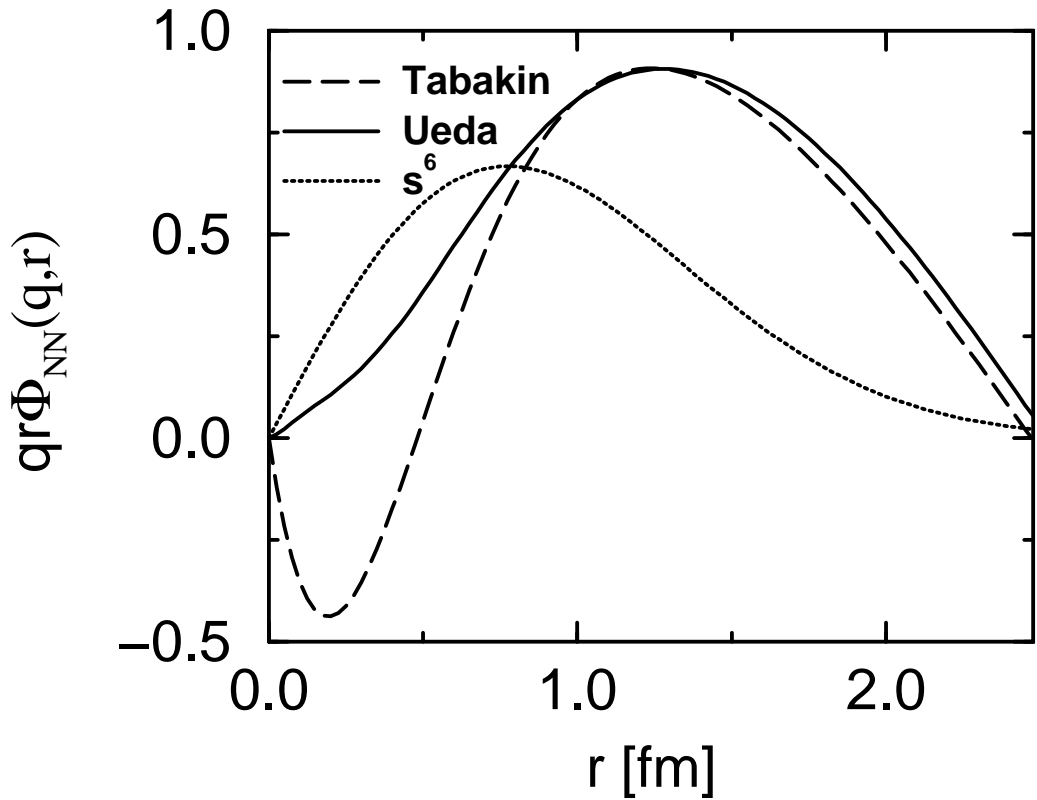}}
\caption[wave]{Wave functions of the final $^1S_0$ state for two NN 
         interaction models [22,23] at fixed lab-energy 
         $E_{NN}$=100 MeV. The projection of the $s^6$ six-quark 
         configuration onto the NN channel is also shown.}
\label{figure:obufig2} 
\end{figure}

We use the non-standard Tabakin potential because 
at short range, the NN wave functions obtained
with this potential differ qualitatively from OBEP wave functions.
In Fig.\ \ref{figure:obufig2}, the wave functions 
(\ref{fst}) and (\ref{fsu}) for both models are
shown at an NN lab-energy of $E_{NN}$=100 MeV. 
The relative wave function of Eq.\ (\ref{fst}) has a node
at distances $r\approx 0.4-0.5$ fm 
(a stable position of the node in a large interval of NN-energies produces 
the same NN-scattering phase shifts as a repulsive core). 
The two models are phase equivalent, but differ in their off-shell behavior. 
In the following we will demonstrate
that the results for $\Gamma_{d^{\prime}}$ differ considerably for
both models, especially if the dibaryon mass $M_{d^{\prime}}$ comes close to 
the $\pi NN$ threshold.


\section{Explicit Calculation using the 
         fractional parentage coefficient (f.p.c.) technique}
\noindent
Our approximation for the decay amplitude in Eq.\ (\ref{tme}) leads to a
sum over products of two factors.
The first factor is the so-called overlap integral  
of the intermediate six-quark configuration with the outgoing two-nucleon
state. The second factor  is a shell-model transition matrix element 
that describes the
production of the pion on a single quark in the dibaryon and the 
subsequent transition
to an intermediate six-quark configuration.
Both factors can be calculated with the standard fractional parentage 
coefficient (f.p.c.) technique, which was developed for quark-model 
calculations for example in Refs.\ \cite{obu79,har,chen,obu82,obu84,so,obu91}.


\subsection{Overlap integral of intermediate six-quark
            configurations with the NN-continuum}
\noindent
In this subsection, we calculate the overlap integral of an 
intermediate six-quark configuration 
$(n,b_6)\,\{f\}$ with the (antisymmetrized and normalized) $^1S_0$ partial 
wave of the final NN-state introduced in Eq.\ (\ref{psinn})
\begin{eqnarray}
&\! & \langle\Psi_{NN}^{L=0}(q)\,|\,(n,b_6)\{f\}\rangle = 
\label{ovrl} \\
&\! & \qquad\; \sqrt{10}\,\langle\Phi_{NN}^{L=0}(q)
3      \{ N((00,b_N)123) N((00,b_N)456)\}_{ST=01} \, |
      \,(n,b_6), [f_X], [f_{CT}] ,LST\!=\!001\rangle \; .
\nonumber
\end{eqnarray}
Beginning with Eq.\ (\ref{ovrl}) we denote from now on the nucleon
wave function of Eq.\ (\ref{n}) of the translationally invariant shell-model
as 
\begin{equation}
\label{nb}
N(123)\equiv N((n^{\prime} l^{\prime}=00,\, b_N)123) \; . 
\end{equation}
Here $l^{\prime}$ is the total orbital angular momentum contained in the 
internal Jacobi coordinates $\bbox{\rho},\bbox{\lambda}$, 
and $b_N$ is the h.o.\ size parameter for the three-quark system.

The overlap integral (\ref{ovrl}) is calculated using the standard f.p.c.\ 
technique for the quark shell model \cite{har,obu82,obu84,so,obu91}. 
For this purpose we use a f.p.c.\ decomposition 
of the six-quark configuration $(n,b_6)\{f\}$ into two three-quark clusters 
with CST quantum numbers of baryons
$\{B_1((n^{\prime} l^{\prime}, b_6)123)B_2((n^{\prime\prime}
l^{\prime\prime}, b_6)456)\}_{CST}$.
Note, that for this procedure, the size parameter in the decomposition $b_6$
differs from the nucleonic size parameter $b_N$
\begin{eqnarray}
&|&(n,b_6)[f_X][f_{CT}]LST\!=\!001\rangle =
\sum_{B_1(n^{\prime})} \sum_{B_2(n^{\prime\prime})}
\sqrt{\frac{n_{f_X^{\prime}}n_{f_X^{\prime\prime}}}{n_{f_X}}} \; 
U_{\{f\}}^{B_1B_2} \cdot C_{f_X}^{(n)}(n^{\prime},n^{\prime\prime})
\nonumber \\
&&\qquad\times\left\{\, \varphi_{\widetilde{N}\widetilde{L}}
(r,\sqrt{2/3}b_6)\, Y_{\widetilde{L}\widetilde{M}}(\hat{\bf r}) \, 
\left\{B_1((n^{\prime} l^{\prime},b_6)123)\, 
B_2((n^{\prime\prime} l^{\prime\prime},b_6)456)\, \right\}_{ST\!=\!01}
\right\}_{L\!=\!0}
\label{fpc}
\end{eqnarray}
In  expansion (\ref{fpc}), $\varphi_{\widetilde{N}\widetilde{L}}(r,\mu r_0)$
is a h.o.\ wave function in the relative coordinate $r$ of the two baryons
with angular momentum
$\widetilde{\vec L}=\vec L- (\vec l^{\prime}+\vec l^{\prime\prime})$ and
$\widetilde{N}     =    n - (n^{\prime}+n^{\prime\prime})$ excitation quanta
($\mu r_0=\sqrt{2/3}b_6$ is the h.o.\ size parameter).
As usual, $n_{f_X}$ is the dimension of the IR
$[f_X]$ of the permutation symmetry group $S_6$ for six particles \cite{ham}.
$n_{f_X^{\prime}}$ and $n_{f_X^{\prime\prime}}$ are the dimensions of IRs
$[f_X^{\prime}]$ and $[f_X^{\prime\prime}]$ of the subgroups
$S_3^{\prime}$ and $S_3^{\prime\prime}$  
in the reduction $S_6\supset S_3^{\prime}\times S_3^{\prime\prime}$.
The coefficients $U_{\{f\}}^{B_1B_2}$ and 
$C_{f_X}^{(n)}(n^{\prime},n^{\prime\prime})$ are f.p.c.\ in the CST- 
and X-subspaces respectively. 
 For simplicity, we omit in Eq.\ (\ref{fpc}) the indices for the dependence of
$U_{\{f\}}^{B_1B_2}$ and $C_{f_X}^{(n)}(n^{\prime},n^{\prime\prime})$
on the intermediate Young schemes 
$f^{\prime}_{CST}\equiv\tilde f^{\prime}_X$,
$f^{\prime\prime}_{CST} \equiv \tilde f^{\prime\prime}_X$, 
$f^{\prime}_{CT}$, $f^{\prime\prime}_{CT}$, $f^{\prime}_S$, 
$f^{\prime\prime}_S$, $f^{\prime}_T$, $f^{\prime\prime}_T$, 
$f^{\prime}_C$, and $f^{\prime\prime}_C$
occuring for our chosen reduction chain of Eq.\ (\ref{su}).

With the help of (\ref{fpc}), we can calculate the overlap (\ref{ovrl}).
The three-quark -- three-quark decomposition 
is of course the most adequate expansion for projecting onto the NN-channel.
The projection for a given intermediate state $(n)\{f\}$
\begin{equation}
\label{phi}
\Phi_{(n)\{f\}}^{L=0}(r)=
\sqrt{10}\langle\{N((00,b_N)123)
N((00,b_N)456)\}_{ST=01}\,|
\,(n,b_6)[f_X][f_{CT}]LST=001\rangle
\end{equation}
receives non-vanishing contributions only from NN-components in 
Eq.\ (\ref{fpc}) because non-nucleonic clusterings, such as  
$B_1(123)B_2(456)$, are orthogonal to $N(123)N(456)$ in CST space.
 Furthermore, the overlap of excited nucleonic clusters,
e.g.\ $N((20,b_6)123)$, with the ground state nucleon $N((00,b_N)123)$ 
can be neglected if we assume that the size parameter $b_6$ of the
six-quark configuration (\ref{fpc}) does not differ considerably from the 
quark core radius $b_N$ of the nucleon. 
In fact, because $b_6\neq b_N$, 
the non-zero overlap integral between  excited and non-excited
nucleons is 
\begin{equation}
  \langle N((20,b_6)123)\,|\,N((00,b_N)123)\rangle =
  \frac{(b_6^2/b_N^2-1)}{(1+b_6^2/b_N^2)}
  \,\left ( \frac{2b_6/b_N}{1+b_6^2/b_N^2}\right )^3 \; .
\label{novrl}
\end{equation}
The sum over all possible terms gives a negligible contribution
to the final result because the different terms interfere destructively 
(see next section).
Due to these restrictions we are lead to the expression
\begin{eqnarray}
\Phi_{(n)\{f\}}^{L=0}(r) &\approx&
  \langle N((00,b_N)123)\,|\,N((00,b_6)123)\rangle\,
  \langle N((00,b_N)456)\,|\,N((00,b_6)456)\rangle\,
\nonumber \\
&\times & \quad
  \sqrt{10} \sqrt{\frac{1}{n_{f_X}}}\, U_{\{f\}}^{NN}\, C_{f_X}^{(n)}(0,0)
  \, \varphi_{n0}(r,\sqrt{2/3}b_6)\, Y_{00}(\hat r)
\label{appr}
\end{eqnarray}
where
$$
  C_{f_X}^{(n)}(0,0)=\left\{
  \begin{array}{cccl}
    1,                  & {\rm{if}} &n=0,&[f_X]=[6]\\
    -\sqrt{\frac{1}{5}},&               &n=2,&[f_X]=[6]\\
    -\sqrt{\frac{4}{5}},&               &n=2,&[f_X]=[42]
  \end{array}
\right.
$$
The coefficients $C_{f_X}^{(n)}(n^{\prime},n^{\prime\prime})$ are calculated
by general methods from the  TISM (see, e.g.\ Ref.\ \cite{nem}).
The values of $U_{\{f\}}^{NN}$  are given in Table \ref{table1}.
The general rule for calculating f.p.c.\ in the CST-subspace is the 
factorization of the value 
$U_{\{f\}}^{B_1B_2}$ \cite{obu79,har,chen,obu82,obu84} (symbolically)  
\begin{equation}
\label{sf}
U_{\{f\}}^{B_1B_2}=SF_{CT\bullet S}\,SF_{C\bullet T}
\end{equation}
in terms of scalar factors $SF_{CT\bullet S}$ and $SF_{C\bullet T}$  
of Clebsch-Gordan coefficients of the unitary groups
$SU(12)_{CST}$ and $SU(6)_{CT}$ for the reductions
$SU(12)_{CST}\supset SU(6)_{CT}\times SU(2)_S$
and $SU(6)_{CT}\supset SU(3)_{C}\times SU(2)_T$, respectively
(which are links of the common reduction chain (\ref{su})). 
The necessary SFs are tabulated in 
Refs.\ \cite{har,chen,obu82,obu84,so,obu91}.
With expression (\ref{appr}), the overlap integral of Eq.\ (\ref{ovrl})
reduces to  
\begin{eqnarray}
\langle\Psi_{NN}^{L=0}(q)\,|\,(n,b_6)\{f\}\rangle &=& 
  \langle\Phi_{NN}^{L=0}(q)\,|\,\Phi_{(n)\{f\}}^{L=0}\rangle
\nonumber \\
&\approx& \sqrt{10}
  \left(\frac{2b_6/b_N}{1+b_6^2/b_N^2} \right)^{\!6} \,
  \sqrt{\frac{1}{n _{f_X}}}\, U_{\{f\}}^{NN}\, C_{f_X}^{(n )}(0,0)
  \left(\frac{ 8\pi b_6^2}{3}\right)^{3/4} I_{NN}^{(n)}(q) \; .
\label{rovrl}
\end{eqnarray}
Eq.\ (\ref{rovrl}) contains a simple radial integral
\begin{equation}
\label{rint}
\left(\frac{8\pi b_6^2}{3}\right)^{3/4}
I_{NN}^{(n)}(q)=\sqrt{4\pi}\int\limits_0^{\infty} r^2\,dr\, 
\Phi_{N N}^{L=0}(q,r) \varphi_{n 0}(r,\sqrt{2/3}b_6) \; ,
\end{equation}
which can be calculated analytically  for a plane wave
$\Phi_{NN}^{L=0}(q,r)=(2\pi)^{-3/2}j_0(qr)$, as well as for the Tabakin 
f.s.i.\ wave functions of Eq.\ (\ref{fst}) and for the Ueda 
f.s.i.\ wave functions given in Eq.\ (\ref{fsu}). 
Results for $I_{NN}^{(n)}(q)$ are listed in the appendix.
The large bracket in Eq.\ (\ref{rovrl}) involving the ratio of the two h.o.\
size parameters $b_N/b_6$ comes from the overlap of the two nucleon clusters 
$\langle N((00,b_N)123)\,|\,N((00,b_6)123)\rangle\,
 \langle N((00,b_N)456)\,|\,N((00,b_6)456)\rangle$.

 
\subsection{Shell-model transition matrix element}
\noindent
The shell-model matrix element of the pion-production operator
$\widehat{\cal{O}}_{\pi q}$ defined in Eq.\ (\ref{o}) 
(the second factor in the decay amplitude introduced in Eq.\ (\ref{tme}))
is proportional to the one-particle  matrix element of the spin-isospin-flip
operator $\sigma_j^{(\mu)}\tau_j^{(\kappa)}$.
The remaining five quarks act as spectators for the transition.
Chosing the sixth quark, we write the matrix element for the emission of a 
$\pi^-$: 
\begin{eqnarray}
\label{smtme}
& & \langle(n,b_6)\{\bar f\}\bar L\bar S\bar T\!=\!001,
     J^P\!\!=0^+,T_z\!=\!1 ;\pi^-          \, | \,
     \widehat{\cal{O}}_{\pi q}({\bf k}) \, | \,
     d^{\prime}\rangle 
\nonumber\\
& & \qquad\qquad =
    6\, \frac{f_{\pi q}}{m_{\pi}}\frac{k}{\sqrt{2E_{\pi}(2\pi)^3}}
    \langle(n,b_6)\{\bar f\}\bar L\bar S\bar T\!=\! 001,T_z\!=\!1\,|\,
    (\bbox{\sigma}_6 \cdot \hat{\bf k})\tau_6^{(+)}
     \exp \left ( {i\frac{5}{6}{\bf k} \cdot \bbox{\rho}_6} \right )\,|\,
    d^{\prime}\rangle 
\end{eqnarray}
Here, $\bbox{\rho}_6={\bf r}_6-\frac{1}{5}\sum_{i=1}^5{\bf r}_i$,
is the Jacobi coordinate, and
$\bbox{\sigma}_6$ and $\bbox{\tau}_6$ are spin and isospin of the sixth quark. 
The momentum of the pion is ${\bf k}=k\hat{\bf k}$, and the factor 6 in front
of the one-particle matrix element contains the summation over all six quarks. 
The natural choice for a f.p.c.\ decomposition is clearly the separation 
of the last quark $q^6\to q^5\times q$ (one-particle f.p.c.), which allows to
exploit the orthogonality constraints for the five spectator quarks.
With the one-particle f.p.c.\
expansion of the shell-model states, the right-hand side of
Eq.\ (\ref{smtme}) reduces to a sum
of one-particle spin-isospin-flip amplitudes with algebraic coefficients:
\begin{eqnarray}
& & \langle(n,b_6)\{\bar f\}\bar L\bar S\bar T\!=\!001,T_z\!=\!1 \,|\,
    (\bbox{\sigma}_6 \cdot \hat{\bf k})\tau_6^{(+)}
    e^{-i\frac{5}{6}{\bf k \cdot \bbox{\rho}_6}} \,|\,
    d^{\prime}\rangle =
    \sum_{\bar f_X^{\prime}} \sqrt{\frac{n_{\bar f_X^{\prime}}}{n_{\bar f_X}}}
    \sum_{     f_X^{\prime}} \sqrt{\frac{n_{     f_X^{\prime}}}{n_{     f_X}}}
\nonumber \\
& & \qquad\times \sum_{\bar S^{\prime},\,\bar T^{\prime}} 
        U_{\{\bar f\}} (\bar S^{\prime}\bar T^{\prime} ,S_6=T_6=1/2)
    \sum_{     S^{\prime},\,     T^{\prime}} 
        U_{\{     f\}} (     S^{\prime}     T^{\prime} ,S_6=T_6=1/2)
    \sum_M (1,-\! M,1,M \,|\, 0,0)
\nonumber \\
& & \qquad\times \left\{ \delta_{n,2}\, C_{[51]_X}^{(1)}(s^4p,s)\,
    \, C_{\bar f_X}^{(2)}(s^4p,p)X_6({\bf k};11M,00)\right.
\nonumber \\
& & \qquad\qquad\qquad\qquad\left. +  C_{[51]_X}^{(1)}(s^5,p) 
    \left[ \delta_{n,0}\, C_{[6]_X}^{(0)}(s^5,s)+
    \delta_{n,2}\, C_{[6]_X}^{(2)}(s^5,2s) \right]X_6({\bf k};11M,n0) 
\right\}
\nonumber \\
& & \qquad\times 3i\langle\bar S^{\prime},S_6:\bar S\!=\!0
    \, | \, (\bbox{\sigma}_6 \cdot \hat{\bf k}) \, | \,
    S^{\prime},S_6:S\!=\!1,M\rangle
    \langle\bar T^{\prime},T_6:\bar T\!=\!T_z\!=\!1
    \, | \, \tau_6^{(+)} \, | \,  T^{\prime},T_6:T\!=\!0\rangle 
\label{series}
\end{eqnarray}
where the functions $X_6({\bf k};11M,n0)$ are spatial integrals
\begin{eqnarray}
& & X_6({\bf k};11M,n0)=\int_{4\pi}(\hat{\bf k} \cdot \hat{\bbox{\rho}}_6)
   Y_{1-M}(\hat{\bbox{\rho}}_6) \, d^2\hat{\bbox{\rho}}_6
\nonumber\\
& & \qquad\qquad\times \int_{0}^{\infty} \,\rho_6^2\,d\rho_6 \,
    \varphi_{n0}(\rho_6,\sqrt{6/5}b_6)
    \varphi_{11}(\rho_6,\sqrt{6/5}b_6)\, j_1(5/6k\rho_6)
\label{spatint}
\end{eqnarray}
Here  $S^{\prime}$ and $T^{\prime}$ ($\bar{S^{\prime}}$ and $\bar{T^{\prime}}$) 
are spin and isospin of the five spectator quarks
after the separation of the sixth quark.
The first term in curly brackets on the right-hand-side of
Eq.\ (\ref{series}) corresponds to the quark transition from an initial
s state to a final p state.
The second term corresponds to the
quark transition from an initial p state to a final s or 2s state.
In Eq.\ (\ref{series}), we use almost the same notation for
one-particle f.p.c.\ $U_{\{f\}}(S^{\prime} T^{\prime},S_6 T_6)$
and $C_{f_X}^{(n)}((n^{\prime}),(n^{\prime\prime}))$ as before for the
three-particle f.p.c.'s in Eq.\ (\ref{fpc}). 
Note, that the one-particle f.p.c.\
$U_{\{f\}}(S^{\prime} T^{\prime},S_6 T_6)$ in the $CST$-subspace can be calculated
with one-particle scalar factors SF (cf.\ Eq.\ (\ref{sf})), which can be found
for example in Ref.\ \cite{so}.

Due to the orthogonality restrictions for the five spectator quarks, 
the summations over
$\bar{f}^{\prime}_X, {f}^{\prime}_X, \bar{S}^{\prime}, \bar{T}^{\prime}, 
 {S}^{\prime}, {T}^{\prime}$ collapse to  
$\delta_{\bar{f}_X^{\prime}, {f}_X^{\prime}}$ 
$\delta_{\bar{S}^{\prime}, {S}^{\prime}}$  $\delta_{\bar{S}^{\prime}, {S}_6}$ 
$\delta_{\bar{T}^{\prime}, {T}^{\prime}}$  $\delta_{\bar{T}^{\prime}, {T}_6}$,
and the only non-vanishing elementary
spin- and isospin-flip amplitudes in Eq.\ (\ref{series}) are 
\begin{eqnarray}
   \langle \bar S^{\prime}=S_6=1/2:\bar S=0 \,| \,\sigma_6^{(\mu)} \, | \,
           S^{\prime}=S_6=1/2:     S=1, -M \rangle 
&=& -(-1)^{\mu}\, \delta_{\mu M}
\nonumber \\
\langle \bar T^{\prime}=T_6=1/2:\bar T=1,\bar T_z \, | \, \tau_6^{(\kappa)}
\, | \, T^{\prime}=T_6=1/2:     T=0 \rangle 
&=& (-1)^{\kappa}\, \delta_{\kappa\bar T_z}
\label{flip}
\end{eqnarray}


\subsection{Decay amplitude after summation over allowed 
            intermediate states}
\noindent
Collecting the shell-model matrix element of Eqs.\ (\ref{smtme})-(\ref{series})
for pion production and the overlap integrals of Eq.\ (\ref{rovrl}) for all
intermediate six-quark configurations, and performing the remaining radial
integrals in Eqs.\ (\ref{rint}) and (\ref{spatint}), 
leads to the following result for the full 
decay amplitude defined in Eq.\ (\ref{tme})
\begin{eqnarray}
  \langle\Psi_{NN}({\bf q}),\pi^- \, | \, \hat {\cal{O}}_{\pi q}({\bf k})
  \, | \, d^{\prime}\rangle  
&=&
  -\frac{10i}{27(2\pi)^{3/2}} \left( \frac{2}{3\pi} \right)^{3/4} \,
  \frac{f_{\pi q}}{m_{\pi}} \sqrt{\frac{b_6}{E_{\pi}}}
  \left( \frac{2b_6/b_N}{1+b_6^2/ b_N^2} \right)^{\!6}
  (kb_6)^2 \exp \left ({-\frac{5}{24}k^2b_6^2} \right )
\nonumber \\
&\times & \left[ I_{NN}^{(0)}(q)+ \sqrt{\frac{2}{27}}
  \left( 1-\frac{k^2b_6^2}{24} \right) I_{NN}^{(2)}(q) \right] \; .
\label{fullam}
\end{eqnarray}
The overlap integrals $I_{NN}^{(0)}(q)$ and $I_{NN}^{(2)}(q)$ 
can be found in the appendix.
Note that the inclusion of overlap terms with excited nucleon configurations in the intermediate six-quark states, originating from the different 
harmonic oscillator parameters $b_6\neq b_N$, given in Eq.\ (\ref{novrl})
leads to a non-essential renormalization factor
of the decay amplitude (\ref{fullam})
\begin{equation}
  1-\frac{1}{5\sqrt{6}}\frac{(b_6^2/b_N^2-1)}{(1+b_6^2/b_N^2)}
  \left (1-\frac{5}{6}k^2b_6^2\right )
\label{renormi}
\end{equation}
in front of the term $I_{NN}^{(0)}(q)$ on the right 
hand side of Eq.\ (\ref{fullam}).
This factor (\ref{renormi}) can be omitted for small $kb_6$.
The $k^2$ behavior of the decay amplitude is due to (i) a factor $k$ in the
transition opertor of eq.(\ref{o}), (ii) due to the fact that a p-wave quark 
is involved either in the initial or
the final state of the one-particle transition matrix element on the
right-hand-side of the Eq.\ (\ref{series}). 
We recall that in the case of the $\Delta$-isobar decay into the 
$\pi N$ channel, the transition matrix element is proportional
only to $k^1$, corresponding to the $\bbox{\sigma}_j \cdot {\bf k}$ term 
in Eq.\ (\ref{o}). 
The $k^2$ behavior of the $d'$ decay amplitude (\ref{fullam}) leads to a very 
strong dependence of the $d^{\prime}$ decay width on the value of 
$M_{d^{\prime}}$, as we will see in the next section.


\section{Numerical results and discussion}
\noindent
The total hadronic decay width of the possible $d^{\prime}$ dibaryon
$\Gamma_{d^{\prime}}$ contains three partial widths 
\begin{equation}
\Gamma_{d^{\prime}}= \Gamma_{\pi^-pp}
                 + \Gamma_{\pi^0pn}
                 + \Gamma_{\pi^+nn} = 3 \Gamma_{\pi^-pp}
\label{gtot}
\end{equation}
which are equal to each other
$\Gamma_{\pi^-pp}=\Gamma_{\pi^0pn}=\Gamma_{\pi^+nn}$, when we neglect isospin
breaking effects. 
The partial $\pi^-pp$ decay width
$\Gamma_{\pi^-pp}$ is defined by the standard expression \cite{iton}
\begin{eqnarray}
\Gamma_{\pi^-pp} &=& 2\pi \int d^3q\int d^3k \,
  \delta\!\left( M_{d^{\prime}}-2M_N-\frac{k^2}{4M_N}-\frac{q^2}{M_N}-
                 \sqrt{m_{\pi}^2+k^2} \right)
\nonumber \\
&\times & \Bigl| \langle\Psi_{NN}({\bf q}),\pi^- \, |
  \, \hat {\cal{O}}_{\pi q}({\bf k}) \, | \, d^{\prime} \rangle \Bigl|^2 \; ,
\label{gampi}
\end{eqnarray}
where  ${\bf q}=\frac{{\bf q}_1-{\bf q}_2}{2}$ is the relative momentum of
the two final protons and ${\bf k}$ is the momentum of emitted pion in
the c.m.\ of the $d^{\prime}$ dibaryon. 
The $\delta$-function conserves the energy in the decay, while the integration
over the momentum conserving $\delta^{(3)}({\bf q}_1+{\bf q}_2+{\bf k})$ 
has already been exploited \cite{iton} in Eq.\ (\ref{gampi}).
The integration over three-particle phase space leads to the following 
result for the partial $d^{\prime}\rightarrow \pi^-pp$ decay width
\begin{eqnarray}
\Gamma_{\pi^-pp} &=& 
  \frac{2^5 10^2\sqrt{6}}{3^8\sqrt{\pi}}
  \frac{f_{\pi q}^2}{4\pi}\frac{1}{m_{\pi}^2}
  \left( \frac{2b_6/b_N}{1+b_6^2/ b_N^2} \right)^{\!12}
\nonumber \\
&\times & \int\limits_0^{q_{max}} 
  \frac{2M_N(k_0b_6)^5}{2M_N+\sqrt{m_{\pi}^2+k_0^2}} \, 
  \exp \left ({-\frac{5}{12}k_0^2b_6^2} \right )
  \left[ I_{NN}^{(0)}(q)+\sqrt{\frac{2}{27}}
  \left( 1-\frac{k_0^2b_6^2}{24} \right) I_{NN}^{(2)}(q) 
  \right]^2 q^2\,dq \; .
\label{result}
\end{eqnarray}
Here, energy conservation relates the pion momentum $k_0$ to the NN
relative momentum $q$ via 
$$
  k_0(q) = \left\{ 4M_N \left[
  \left( M_{d^{\prime}} - \frac{q^2}{M_N} \right) -
  \sqrt{ \left( M_{d^{\prime}} - \frac{q^2}{M_N} \right)^2 -
  \left( M_{d^{\prime}} -2M_N - \frac{q^2}{M_N} \right)^2+m_{\pi}^2 }
  \right] \right\}^{1/2}
$$ 
and for $q_{max} = \sqrt{M_N(M_{d^{\prime}}-2M_N-m_{\pi})}$, all available
decay energy is converted to kinetic energy in the relative NN-system, and none
to the pion $E_\pi = m_\pi, k_0=0$.

The calculated decay widths are shown in Table \ref{table2}, 
where we have introduced the 
abbreviations p.w., T and U. Here,  p.w.\ refers to a calculation employing
a plane-wave final N-N state (\ref{pw}), while T and U refer to 
calculations using the Tabakin \cite{tab} (T)  and Ueda et al.\ \cite{ued} (U) 
separable NN potentials for the final state interaction.

In parentheses we give the results obtained in the approximation of 
using only one intermediate six-quark configuration $s^6$ (n=0).
With the exception of the results for the Ueda NN-potential 
for sets 1, 2 and 4 (for which the $d^{\prime}$ mass is 400
-- 650 MeV above the $\pi NN$ threshold), 
the inclusion of all Pauli principle allowed intermediate
2$\hbar\omega$ shell-model configurations tends to increase the decay width
by some 20 -- 30 $\%$. 
The largest effect is obtained for $d^{\prime}$ masses rather close to threshold,
exemplarily shown for sets 3 and 5. 

\begin{figure}[h]
  {\epsffile{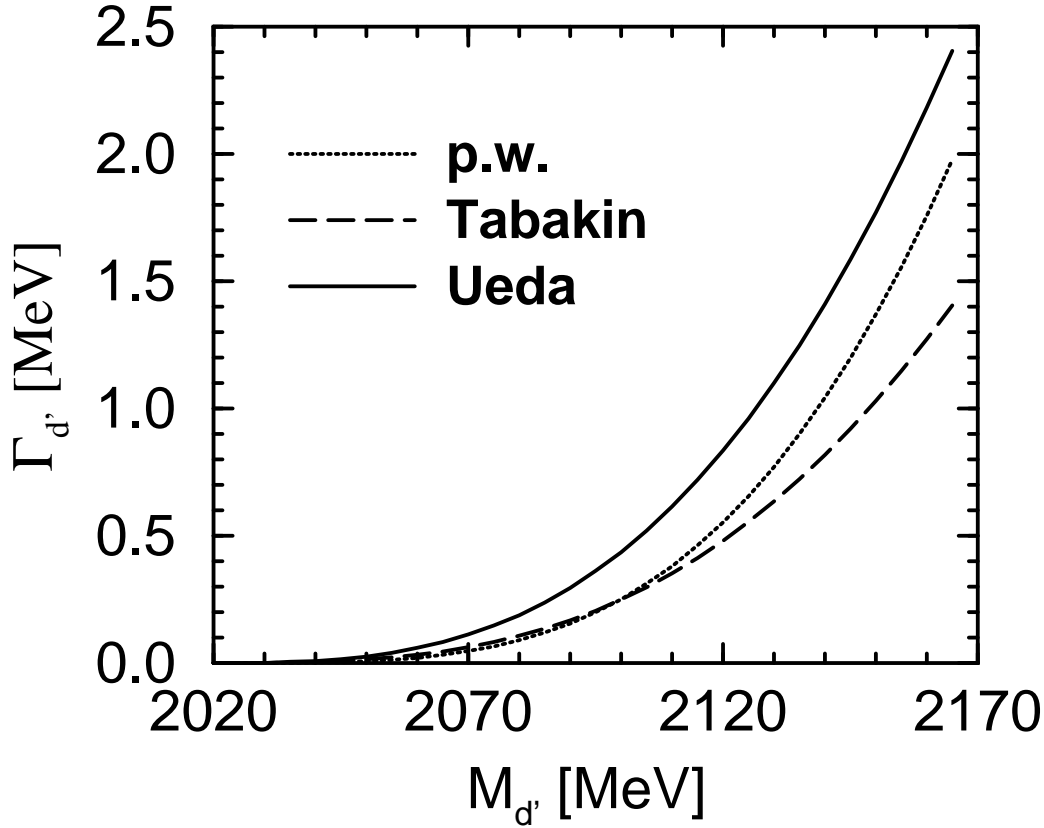}}
\caption[pionic]{Pionic decay width $\Gamma_{d^{\prime}}$ 
         of the $d^{\prime}$ as a function of the dibaryon mass 
         $M_{d^{\prime}}$ for different final state interactions (f.s.i.) 
         between the outgoing nucleons:
         (i) plane wave (p.w., dotted curve), 
         (ii) with f.s.i. using the Tabakin [22] potential 
         (dashed curve), 
         (iii) with f.s.i. using the Ueda {\it et al.} [23] potential 
         (plain curve). 
         The harmonic oscillator parameters $b_N$=0.595 fm and $b_6$=0.95 fm 
         are those of set 5 in Table 2.
         The $\pi$NN threshold for the decay is at 2016 MeV, while the 
         experimentally suggested resonance position of the $d^{\prime}$ is
         at 2065 MeV.} 
\label{figure:obufig3} 
\end{figure}

It can be seen from Table \ref{table2} and Fig.\ \ref{figure:obufig3},
that the pionic decay width of 
the $d^{\prime}$ is very sensitive to the dibaryon mass $M_{d^{\prime}}$, 
which determines the available phase space of the three-body $\pi NN$ decay. 
The sensitivity grows dramatically near the $\pi NN$ threshold (2016 MeV). 
If we extrapolate the results of Table \ref{table2} to the experimental value 
of $M_{d^{\prime}}$ = 2065 MeV, we obtain a very strong reduction
of $\Gamma_{d^{\prime}}$ as compared with the quite realistic variants 
(sets 3 and 5) in Table \ref{table2}:
$$
  \begin{array}{cccl}
    \Gamma_{d^{\prime}}^{p.w.} = 0.032\,{\rm MeV}, &
    \Gamma_{d^{\prime}}^{T}    = 0.046\,{\rm MeV}, &
    \Gamma_{d^{\prime}}^{U}    = 0.083\,{\rm MeV}, & 
    \; \mbox{if} \; b_N \mbox{=0.595 fm and}\;  b_6 \mbox{=0.95 fm} \\
    \Gamma_{d^{\prime}}^{p.w.} = 0.018\,{\rm MeV}, &
    \Gamma_{d^{\prime}}^{T}    = 0.045\,{\rm MeV}, &
    \Gamma_{d^{\prime}}^{U}    = 0.040\,{\rm MeV}, &
    \; \mbox{if} \; b_N \mbox{=0.6 fm and}\;  b_6 \mbox{=1.24 fm}\; .
  \end{array}
$$
This strong dependence of $\Gamma_{d^{\prime}}$ on 
the value of $M_{d^{\prime}}$
is a consequence of the high power of $(k_0b_6)^5$ in the
integrand of Eq.\ (\ref{result}). 
The origin of this $k_0^5$ behavior  
(compared with a $k_0^3$ behavior in case of the $\Delta$-isobar decay) 
comes, as explained above, from the necessity to excite (or de-excite) 
a p-wave quark for the production of a pion.
Note that for small $q_{max}$ (when $M_{d^{\prime}}$ is close to the $\pi$NN 
threshold) the function $k_0(q)$ is linear in the 
factor $\sqrt{q^2_{max}-q^2}$ and can be written as 
$k_0(q)\approx q_{max}\sqrt{{4m_\pi} (1-q^2/q_{max}^2)/M_{d^{\prime}}}$.
Therefore, for small $q_{max}$ the integral in Eq.\ (\ref{result})  
behaves as $q_{max}^8$.
The second high-power factor in Eq.\ (\ref{result}) is the scale factor
$$
\left( \frac{2 b_6/b_N}{1+b_6^2/b_N^2} \right)^{\! 12},
$$ 
which depends sensitively on the ratio $b_6/b_N$. 
However, this sensitivity is considerably reduced by the factor $b_6^5$ 
in the integrand.
The product 
$$b_6^5 \left( \frac{2 b_6/b_N}{1+b_6^2/b_N^2} \right)^{\! 12}$$ 
is a quite smooth function of $b_N/b_6$.
For $b_N$=0.6 fm this product varies from 0.078 fm$^5$ to 0.158 fm$^5$,
if $b_6$ varies from 0.6 fm to 1.24 fm.

For small $q_{max}$, f.s.i.\ make an important contribution 
 to the $d^{\prime}$ decay width because of the large scattering length in the
$^1S_0$ wave $a_s=-23.7$ fm. 
The f.s.i. enhances the decay width for example
by about $85\%$ for set 5 in Table \ref{table2}. 
At the experimental mass $M_{d^{\prime}}$=2065 MeV, the hadronic decay width is
more than doubled by the final state interaction.
It is interesting that in the case of the Tabakin model with a nodal
NN wave function at short range, the contribution from f.s.i.\ is smaller
than for the Ueda model and can even decrease the width compared to the plane
wave result (cf.\ set 4). 
This is a direct consequence of an approximate
orthogonality of the nodal wave function of the Tabakin model to the
projection of the intermediate $s^6$ configuration  
(i.e.\ the h.o.\ function $\varphi_{00}$) 
of Eq.\ (\ref{appr}) onto the NN channel. 
This can easily be seen from Fig.\ \ref{figure:obufig2}, 
where both wave functions are shown. 
The approximate orthogonality of the functions
$\varphi_{00}$ and $\Phi_{NN}^{Tabakin}$ in the integrand of Eq. (29)
reduces considerably the overlap factor $I_{NN}^{0}(q)$,
 which gives the dominant contribution to the $d^{\prime}$ decay width 
(see values in parenthesis in Table \ref{table2}).
As it can be seen in Fig.\ \ref{figure:obufig3}, 
the disagreement between the Tabakin and Ueda models grows with
increasing dibaryon mass $M_{d^{\prime}}$ 
(the influence of the large scattering length $a_s$, which is common for both 
 models, becomes negligible compared to the effect of the larger phase space). 
 For sets 1, 2 and 4 in Table \ref{table2},
the Tabakin model leads again to values of $\Gamma_{d^{\prime}}$,
which are even smaller than $\Gamma_{d^{\prime}}$ in the plane wave 
approximation neglecting f.s.i.


\section{Summary}
\noindent
In this work we have studied the pionic decay of a possible
$d^{\prime}$ dibaryon within the microscopic quark shell-model.
We use a single-quark transition operator which describes the production of the
pion on a single quark.
The dibaryon wave function is given as a single six-quark
translationally invariant shell model configuration, which has been found to
provide an adequate description of the $d^{\prime}$ \cite{wag95,buch95}. 
Previous results from \cite{iton} have been improved
mainly in three points, leading to a complete calculation in the sense, that
i) the  calculation is performed consistently
   on the quark level, i.e.\ the final 
   two-nucleon state is normalized and antisymmetrized on the quark level,
ii) all important intermediate six-quark states with nonvanishing overlaps
   with the final two nucleons are included, and
iii) the strong final state interaction for the two nucleons is
   taken into account on the basis of the separable 
   Tabakin \cite{tab} and Ueda {\sl et al. } \cite{ued} NN potentials. 

Not surprisingly, the small available phase space in the three-body decay
is the dominant mechanism for the narrow width of the $d^{\prime}$. 
The large identity factor (15), on the other hand, enhances the
results of previous evaluations disregarding the 
identity of quarks from different nucleons.
The inclusion of all Pauli principle allowed intermediate
2$\hbar\omega$ shell-model configurations tends to increase the decay width
by some 20 -- 30 $\%$. 
 Furthermore, final state interaction for the two outgoing nucleons also
increases the decay width considerably, if the $d^{\prime}$ mass is close to the
$\pi NN$ threshold.  
Due to these three effects, the calculated pionic decay widths lie between
$\Gamma_{d^{\prime}}$ = 0.18--0.32 MeV for the  most realistic 
set 5, having a $d^{\prime}$ mass close to the experimentally suggested one
and a characteristic hadronic size of the dibaryon of $b_6\approx$ 1 fm. 
This qualitatively agrees with the experimentally suggested value
$\Gamma_{d^{\prime}}$ = 0.5 MeV.            

Despite the fact that both, the 
Tabakin and Ueda f.s.i.\ models,  are unsuitable for 
large NN energies (as in parameter sets 1, 2 and 4), 
the two models demonstrate the strong influence of the
short-range behavior of the NN wave function on the $d^{\prime}$ decay width. 
 (See e.\ g.\ Figs.\  \ref{figure:obufig2} and \ref{figure:obufig3}). 
Recall that these two models are typical representatives
of qualitatively different classes of NN phenomenology. Whereas the Ueda
separable potential is an approximation of the OBEP, 
i.\ e.\ a model with short-range repulsion,
the Tabakin potential can be considered as a unitary-pole approximation (UPA)
\cite{nakai} of a Moscow-type potential model \cite{kuku} with short-range 
attraction and forbidden states. The Moscow model proceeds from the assumption
of a six-quark origin of the short-range NN interaction and pretends to give
an adequate description of the non-local character of the NN force.
The main conclusion to be drawn here is, that these two models, 
which are phase equivalent, differ considerably in their effect on 
the $d^{\prime}$ decay width.
Therefore, a possible $d^{\prime}$ dibaryon would provide a natural laboratory
for detailed studies of the short-range NN interaction.

An interesting continuation of this work would be to go 
beyond phenomenological NN potential models and use a completely 
microscopic quark model approach 
(see e.g.\ \cite{faess} and references therein).
For example, one could calculate the pionic 
decay of the $d^{\prime}$ dibaryon using a final $^1\! S_0$ NN-scattering
wave function that is based on the same microscopic quark
Hamiltonian which simultaneously describes the mass $M_{d^{\prime}}$ and 
structure of the $d^{\prime}$ dibaryon. However, such a calculation is 
complicated by the fact that we have used two different Hamiltonians,
i.\ e.\  two different confinement strengths, 
for three-quark baryons and the six-quark $d^{\prime}$ 
dibaryon \cite{buch95}. Thus, the $d^{\prime}$ 
dibaryon could not be explained in terms of the standard
constituent quark model, using a common Hamiltonian for any number of
quarks. On the other hand, if the $d'$ really exists, this may be taken 
as an indication that the effective (nonperturbative) 
quark-quark interaction depends on the state of the system.


\appendix
\section{}
\noindent
In this appendix we present the analytical expressions for the radial 
integrals $I_{NN}^{(n)}(q)$ defined in Eq.\ (\ref{rint}), which
are needed to calculate the overlap integral of Eq.\ (\ref{ovrl}) between
different intermediate six-quark shell-model configurations and the two
outgoing nucleons. 
We recall, that the relative NN-wave function may be described by a simple
plane wave (p.w.) $\Phi_{NN}^{L=0}(q,r)=(2\pi)^{-3/2}j_0(qr)$ 
or by f.s.i.\ wave functions resulting, for example,
from  separable potential representations of the NN-interaction.
We use the results for the NN-wave function obtained by Tabakin \cite{tab}
given in Eq.\ (\ref{fst}) and the result obtained by Ueda and co-workers
\cite{ued} given in Eq.\ (\ref{fsu}).
Note that coefficients $\tilde A$ and $\tilde B_n$ in Eq. (\ref{fsu})
depend on parameters of the one-term separable potential of Ueda et al.
\cite{ued} $V(q,q^{\prime})=-M_{11}g(q)g(q^{\prime})$, 
$g(q)=\sum_n\frac{c_n(q_c^2-q^2)}{(q^2+\beta_n^2)(q^2+\gamma^2)}$.
For $\tilde A$ and $\tilde B_n$ we use the following expressions:
$$
\tilde A(q)=\alpha^2\frac{g(q)}{1-G(q)}\sum_n\frac{c_n}{\gamma^2-\beta_n^2}
\left (\frac{\gamma^2+q_c^2}{\gamma^2+q^2}\right)\,,
$$

$$
\tilde B_n(q)=\alpha^2\frac{g(q)}{1-G(q)}
\left (\frac{c_n}{\gamma^2-\beta_n^2}\right)
\left (\frac{\beta_n^2+q_c^2}{\beta_n^2+q^2}\right)\,,
$$

\noindent where $\alpha^2=2\pi^2m_NM_{11}/\hbar^2$, 
$G(q)=\frac{2}{\pi}\alpha^2
\int_0^{\infty}[(g^2(k)k^2-g^2(q)q^2)/(k^2-q^2)]dk$
and a value of $\gamma$, which is not fixed in the
ref.~\cite{ued}, is fitted to the singlet scattering length 
($a_s$=-23.7 fm), $\gamma=11.114 \,\,$fm$^{-1}$. 

In Table \ref{table:appendix} we introduced the following abbreviations:
\begin{eqnarray*} 
f^{(0)}(x) &=& e^{-x^2/3},
\\
f^{(2)}(x) &=& -\sqrt{\frac{3}{2}} \left(1-\frac{4}{9}x^2\right)e^{-x^2/3},
\\
g^{(0)}(x) &=& \frac{\sqrt{3}}{x\sqrt{\pi}}-e^{-x^2/3}
               Im\Phi \left(\frac{ix}{\sqrt{3}}\right),
\\
g^{(2)}(x) &=& -\sqrt{\frac{3}{2}}\left[\frac{\sqrt{3}}{x\sqrt{\pi}}
 \left(\frac{1}{3}- \frac{4}{9}x^2\right)-
 \left(      1    - \frac{4}{9}x^2\right) e^{-x^2/3}
 Im\Phi \left( \frac{ix}{\sqrt{3}}\right) \right],
\\
F^{(0)}(x) &=& \frac{\sqrt{3}}{x\sqrt{\pi}}-e^{x^2/3} 
 \left[ 1-\Phi(x/3) \right],
\\
F^{(2)}(x) &=& -\sqrt{\frac{3}{2}} \left\{\frac{\sqrt{3}}{x\sqrt{\pi}}
 \left(\frac{1}{3}+\frac{4}{9}x^2\right) - \left(1+\frac{4}{9}x^2\right)
 e^{x^2/3}\left[1-\Phi(x/3)\right]\right\},
\\
G_1^{(0)} &=& \frac{\sqrt{3}}{x\sqrt{\pi}}+ \left( \sin\,x^2/3- 
 \cos\,x^2/3 \right) \left[ 1-Re\Phi \left((1+i)x/\sqrt{3} \right) \right]-
\\
& & \left( \sin\,x^2/3+ \cos\,x^2/3\right) Im\Phi\left((1+i)x/\sqrt{3}\right),
\\
G_2^{(0)} &=& \left( \sin\,x^2/3+ \cos\,x^2/3\right)
 \left[1-Re\Phi\left((1+i)x/\sqrt{3}\right)\right]+
\\
& & \left( \sin\,x^2/3- \cos\,x^2/3\right) Im\Phi\left((1+i)x/\sqrt{3}\right),
\\
G_1^{(2)} &=& -\sqrt{\frac{3}{2}}\left\{ \frac{\sqrt{3}}{x\sqrt{\pi}}+
 \left[\left(\frac{1}{3}+\frac{4}{9}x^2\right) \sin\,x^2/3-
       \left(\frac{1}{3}-\frac{4}{9}x^2\right) \cos\,x^2/3\right]
 \left[1-Re\Phi \left((1+i)x/\sqrt{3}\right) \right] \right.
\\
& - & \left.\left[ \left( \frac{1}{3}-\frac{4}{9}x^2\right) \sin\,x^2/3 +
                   \left( \frac{1}{3}+\frac{4}{9}x^2\right) \cos\,x^2/3 \right]
 Im\Phi \left( (1+i)x/\sqrt{3} \right) \right\},
\\
G_2^{(2)} &=& -\sqrt{\frac{3}{2}}
 \left\{ -\frac{4}{9} x^2 \frac{\sqrt{3}}{x\sqrt{\pi}} +
 \left[ \left( \frac{1}{3}-\frac{4}{9}x^2 \right) \sin\, x^2/3 +
        \left( \frac{1}{3}-\frac{4}{9}x^2 \right) \cos\, x^2/3 \right]
 \left[ 1-Re\Phi \left( (1+i)x/\sqrt{3} \right) \right] \right.
\\
& - & \left. \left[ \left( \frac{1}{3}+\frac{4}{9} x^2 \right) \sin\,x^2/3 -
                    \left( \frac{1}{3}-\frac{4}{9} x^2 \right) \cos\,x^2/3 
 \right] Im\Phi \left( (1+i)x/\sqrt{3} \right) \right\}
\end{eqnarray*}


\noindent
{\bf Acknowledgments:}
\\
\noindent
Two of the authors (K.I. and I.T.O.) acknowledge the warm hospitality
extended to them at the Institute for Theoretical
Physics, University of T\"{u}bingen, where this work has been mostly completed. 
One of the authors (K.I.) wishes to thank the Ministry
of Education, Science, Sports and Culture (Japan) for financial support.
Partially, this work was supported with the RFBR grant No 96-02-1807 and one 
of the authors (I.T.O.) thanks RFBR for financial support.
G.W.\ thanks the Deutsche Forschungsgemeinschaft (DFG) for a postdoctoral 
 fellowship, contract number WA1147/1-1.


\newpage
{\Large I.\ T.\ Obukhovsky {\sl et al.}, {\bf Figure 1}, 

\vspace{.2cm}
"Pionic decay of a possible $d^{\prime}$ dibaryon and ...''} 

\vspace{2.2cm}

\newpage
{\Large I.\ T.\ Obukhovsky {\sl et al.}, {\bf Figure 2}, 

\vspace{.2cm}
"Pionic decay of a possible $d^{\prime}$ dibaryon and ...''} 

\vspace{2.2cm}

\newpage
{\Large I.\ T.\ Obukhovsky {\sl et al.}, {\bf Figure 3}, 

\vspace{.2cm}
"Pionic decay of a possible $d^{\prime}$ dibaryon and ...''} 

\vspace{2.2cm}

\newpage
\noindent
{\bf\underline{Tables:}}

\begin{table}[h]
\begin{tabular}{c|c|ccccc} \hline 
$[f_{CTS}]$ & $[1^6]_{CTS}$ &  &  &   $[2^21^2]_{CTS}$  &  & 
\\ [0.15cm] \hline
$[f_{CT}]$ & $[2^3]_{CT}$ & $[42]_{CT}$ & $[321]_{CT}$ & $[2^3]_{CT}$ &
$[31^3]_{CT}$ & $[21^4]_{CT}$
\\[0.15cm] \hline\hline
$U_{\{f\}}^{NN}$ & $\sqrt{\frac{1}{9}}$ & $-\sqrt{\frac{9}{20}}$ & 
$\sqrt{\frac{16}{45}}$ & $\sqrt{\frac{1}{36}}$ & $-\sqrt{\frac{1}{18}}$ & 0
\\ [0.15cm] \hline 
\end{tabular}
  \caption[Cst]{The CST part of the f.p.c.\ three-quark -- three-quark 
           decomposition for the projection onto the NN-channel 
           $U_{\{f\}}^{NN}$, $\{f\}=\{[f_{CTS}],[f_{CT}]\}$}
  \label{table1}
\end{table}

\begin{table}[h]
\begin{tabular}{ c | c  c  c | c  c  c | c  c  c }
\hline
 & & & & & $\Gamma_{\pi^-pp}$[MeV] & & & $\Gamma_{d^{\prime}}$[MeV] & 
\\ [0.15cm] \hline
    set & $b_N$[fm] & $b_6[fm]$ & $M_{d^{\prime}}$[MeV] & 
    p.w.\ & T & U & p.w.\ & T & U
\\ [0.15cm] \hline
1 & 0.45  & 0.59 & 2705 & 56.8(44.1)   & 7.8(5.2)     & 41.2(46.0)   & 170.5 
  & 23.3  & 123.6 \\
2 & 0.47  & 0.65 & 2680 & 44.2(35.7)   & 8.3(7.0)     & 32.6(36.5)   & 132.5 
  & 24.9  & 97.8  \\
3 & 0.6   & 1.24 & 2162 & 0.22(0.17)   & 0.27(0.18)   & 0.25(0.22)   & 0.67  
  & 0.81  & 0.76  \\
4 & 0.595 & 0.78 & 2484 & 28.3(22.6)   & 10.9(8.9)    & 21.8(21.9)   & 84.8  
  & 32.6  & 65.4  \\
5 & 0.595 & 0.95 & 2092 & 0.058(0.036) & 0.061(0.049) & 0.107(0.071) & 0.173 
  & 0.183 & 0.321 
\\ [0.15cm] \hline
\end{tabular}
  \caption[pion decay]{Calculated 
           $\pi^-$ decay width $\Gamma_{\pi^-pp}$ and the total 
           hadronic decay width $\Gamma_{d^{\prime}}$ of the $d^{\prime}$ 
           dibaryon for five different $d^{\prime}$ masses and wave 
           functions ($b_6$ is the characteristic $d^{\prime}$ size 
           parameter, and $b_N$ is the quark core radius of the nucleon). 
           Masses and wave functions of the $d^{\prime}$ were obtained in 
           Refs.\ [6,7] within different models for 
           the microscopic q-q interaction.}
  \label{table2}
\end{table}

\begin{table}[h]
\begin{tabular}{c|c|c} \hline 
Model & $I_{NN}^{(0)}(q)=$ & $I_{NN}^{(2)}(q)=$
\\ [0.15cm] \hline
p.w. & $f^{(0)}(qb_6)$ & $f^{(2)}(qb_6)$
\\ [0.15cm] \hline
Tabakin & $cos\delta_0 \left[ f^{(0)}(qb_6)+tan\delta_0g^{(0)}(qb_6)+ \right. $
        & $cos\delta_0 \left[ f^{(2)}(qb_6)+tan\delta_0g^{(2)}(qb_6)+ \right. $
\\
Eq.\ (\ref{fst}) & $A(q)\alpha F^{(0)}(\alpha b_6)+
                    B_1(q)\beta G^{(0)}_1(\beta b_6) +$   
                 & $A(q)\alpha F^{(2)}(\alpha b_6)+ 
                    B_1(q)\beta G^{(2)}_1(\beta b_6) +$ 
\\
 & $\left. B_2(q)\beta G^{(0)}_2(\beta b_6) \right] $  
 & $\left. B_2(q)\beta G^{(2)}_2(\beta b_6) \right] $
\\ [0.15cm] \hline
Ueda & $cos\delta_0 \left[ f^{(0)}(qb_6) + tan\delta_0g^{(0)}(qb_6)+\right. $
     & $cos\delta_0 \left[ f^{(2)}(qb_6) + tan\delta_0g^{(2)}(qb_6)+\right. $
\\
Eq.\ (\ref{fsu}) & $\left. \widetilde{A}(q)\gamma F^{(0)}(\gamma b_6) - 
                \sum_n\widetilde{B}_n(q)\beta_n F^{(0)}(\beta_n b_6)\right]$ 
                 & $\left. \widetilde{A}(q)\gamma F^{(2)}(\gamma b_6) -
                \sum_n\widetilde{B}_n(q)\beta_n F^{(2)}(\beta_n b_6)\right]$
\\ [0.15cm] \hline
\end{tabular}
  \caption[radial]{Radial integrals $I_{NN}^{(n)}(q)$ for plane waves (p.w.) and 
           f.s.i.\ functions given in Eqs.\ (20) and (21) 
           for the Tabakin [22] and Ueda [23] separable 
           potential formulation of the relative NN-wave function.}
  \label{table:appendix}
\end{table}


\begin{thebibliography}{99}
\bibitem{bil91} 
  R.\ Bilger, B.\ M.\ Barnet, H.\ Clement, S.\ Krell, 
  G.\ J.\ Wagner, J.\ Jaki, C.\ Joram, T.\ Kirchner, W.\ Kluge, M.\ Metzler, 
  R.\ Wieser, and D.\ Renker, Phys.\ Lett.\ B {\bf 269}, 247 (1991); 
  R.\ Bilger, H.\ Clement, K.\ F\"ohl, K.\ Heitlinger, C.\ Joram,
  W.\ Kluge, M.\ Schepkin, G.\ J.\ Wagner, R.\ Wieser, R.\ Abela, F.\ Foroughi,
  and D.\ Renker, Z.\ Phys.\ A {\bf 343}, 941 (1992);
  K. F\"ohl, Thesis, University of Tuebingen 1996 (submitted for publication);
  H.\ Clement, M.\ Schepkin, G.\ J.\ Wagner, and O.\ Zaboronsky, 
  Phys.\ Lett.\ B {\bf 337}, 43 (1994).
\bibitem{kam}
  W.\ A.\ Kaminski, Phys.\ Part.\ Nucl.\ {\bf 26}, 148 (1995).
\bibitem{mart91}
  B.\ V.\ Martemyanov and M.\ G.\ Schepkin, 
  JETP Lett.\ {\bf 53}, 776 (1991);
  R.\ Bilger, H.\ A.\ Clement, and M.\ G.\ Schepkin, 
  Phys.\ Rev.\ Lett.\ {\bf 71}, 42 (1993);
  R.\ Bilger, $\pi N$ Newsletter {\bf 10}, 47 (1995).
\bibitem{Clem95} 
  H. Clement (private communication).
\bibitem{muld}
  P.\ J.\ Mulders, A.\ T.\ Aerts, and J.\ J.\ de Swart, 
  Phys.\ Rev.\ Lett.\ {\bf 40}, 1543 (1978); 
  Phys.\ Rev.\ D {\bf 21}, 2653 (1980).
\bibitem{wag95}
  Georg Wagner, L.\ Ya.\ Glozman, A.\ J.\ Buchmann, 
  and Amand Faessler, Nucl.\ Phys.\ {\bf A594}, 263 (1995); 
  L.\ Ya.\ Glozman, A.\ J.\ Buchmann, and Amand Faessler, 
  J.\ Phys.\ G {\bf 20}, L49 (1994).
\bibitem{buch95}
  A.\ J.\ Buchmann, Georg Wagner, and Amand Faessler, 
  submitted to Phys.\ Rev.\ C (1997);  
  A.\ J.\ Buchmann, Georg Wagner, K.\ Tsushima, Amand Faessler, and
  L.\ Ya.\ Glozman, $\pi N$ Newsletter {\bf 10}, 68 (1995);  
  Prog.\ Part.\ Nucl.\ Phys.\ {\bf 36}, 383 (1996).
\bibitem{iton}
  K.\ Itonaga, A.\ J.\ Buchmann, Georg Wagner, and Amand Faessler,
  Nucl.\ Phys.\ {\bf A609}, 422 (1996).
\bibitem{seth}
  K.\ K.\ Seth, Proc.\ Int.\ {\it Pions in Nuclei}, edited by E.\ Oset
  (World Scientific, Singapore, 1992) p.\ 205.
\bibitem{gold95}
  F.\ Wang, J.-L.\ Ping, G.-H.\ Wu, L.-J.\ Teng, and T.\ Goldman,
  Phys.\ Rev.\ C {\bf 51}, 3411 (1995).
\bibitem{gar}
  B.\ Schwesinger and N.\ N.\ Scoccola, 
  Phys.\ Lett.\ B {\bf 363}, 29 (1995); 
  H.\ Garcilazo and L.\ Mathelisch, 
  Phys.\ Rev.\ Lett.\ {\bf 72}, 2971 (1994); 
  A.\ Valcarce, H.\ Garcilazo, F.\ Fernandez, and E.\ Moro, 
  Phys.\ Rev.\ C {\bf 54}, 1010 (1996); 2085 (1996). 
\bibitem{sim}
  F.\ \u{S}imkovic and A.\ Faessler, 
  Few-Body Systems Suppl.\ {\bf 9}, 231 (1996).
\bibitem{schep}
  M.\ Schepkin, Preprint of Uppsala Univ.\ TSL/ISV-96-01 41 (1996).
\bibitem{obu79}
  I.\ T.\ Obukhovsky, V.\ G.\ Neudatchin, Yu.\ F.\ Smirnov, and 
  Yu.\ M.\ Tchuvil'sky, Phys.\ Lett.\ B {\bf 88}, 231 (1979).
\bibitem{har}  
  M.\ Harvey, Nucl.\ Phys.\ {\bf A352}, 301 (1981); 326 (1981).
\bibitem{chen}
  Jin-Quan\ Chen, J.\ Math.\ Phys.\ {\bf 22}, 1 (1981);
  J.\ Q.\ Chen, {\it Group Representation Theory for Physicist} 
  (World Scientific, Singapore, 1989);
  J.\ Q.\ Chen {\it et al.}, {\it Tables of the} 
  $SU(mn)\subset SU(m)\times SU(n)$
  {\it Coefficients of Fractional Parentage} (World Scientific,
  Singapore, 1991).
\bibitem{obu82}
  I.\ T.\ Obukhovsky, Yu.\ F.\ Smirnov, and Yu.\ F.\ Tchuvil'sky, 
  J.\ Phys.\ A {\bf 15}, 7 (1982);
  I.\ T.\ Obukhovsky, Z.\ Phys.\ A {\bf 308}, 253 (1982);
  Prog.\ Part.\ Nucl.\ Phys.\ {\bf 36}, 359 (1996).
\bibitem{obu84}
  V.\ G.\ Neudatchin, I.\ T.\ Obukhovsky, and Yu.\ F.\ Smirnov, 
  Phys.\ Part.\ Nucl.\ {\bf 15}, 519 (1984).
\bibitem{so}
  S.\ T.\ So and D.\ Strotmann, J.\ Math.\ Phys.\ {\bf 20}, 153 (1979).
\bibitem{obu91}
  A.\ M.\ Kusainov, V.\ G.\ Neudatchin, and I.\ T.\ Obukhovsky, 
  Phys.\ Rev.\ C {\bf 44}, 2343 (1991).
\bibitem{fa82}
  A.\ Faessler, F.\ Fernandez, G.\ L\"ubeck, and K.\ Shimizu,
  Phys.\ Lett.\ B {\bf 112}, 201 (1982);
  Nucl.\ Phys.\ {\bf A402}, 555 (1983).
\bibitem{tab}
  F.\ Tabakin, Phys.\ Rev.\ {\bf 174}, 1208 (1968).
\bibitem{ued}
  T.\ Ueda, K.\ Tada, and T.\ Kameyama, 
  Prog.\ Theor.\ Phys.\ {\bf 95}, 115 (1996).
\bibitem{nem}
  O.\ F.\ Nemetz, V.\ G.\ Neudatchin, A.\ T.\ Rudchik, Yu.\ F.\ Smirnov, and
  Yu.\ F.\ Tchuvil'sky, 
  {\it Nucleon Clusters in Atomic Nuclei and Many-Nucleon Transfer Reactions}, 
  Monograph in Russian (Kiev, Naukova Dumka, 1988); 
  V.\ G.\ Neudatchin, Yu.\ F.\ Smirnov, and N.\ F.\ Golovanova, 
  Adv.\ Nucl.\ Phys.\ {\bf 11}, 1 (1978).
\bibitem{ham}
  M.\ Hamermesh, {\it Group Theory and its Application to Physical Problems}
  (Adison-Wesley, Massachusetts-Palo Alto-London, 1964).
\bibitem{nakai}
  S.\ Nakaishi-Maeda, Phys.\ Rev.\ C {\bf 51},1663 (1995). 
\bibitem{kuku}
  V.\ I.\ Kukulin, V.\ M.\ Krasnopol'sky, V.\ N.\ Pomerantzev, and B.\ P.\
  Sazonov, Phys.\ Lett.\ B {\bf 153}, 7 (1985).
\bibitem{faess}
  A.\ Faessler, A.\ Buchmann, and Y.\ Yamauchi, Int.\ J.\ Mod.\ Phys.\ 
  {\bf 2}(1), 39 (1993). 
\end{thebibliography}
\end{document}